\documentclass[reprint,showpacs,showkeys,amsmath,amssymb,aps,nofootinbib]{revtex4-1}

\usepackage{graphicx}% Include figure files
\usepackage{dcolumn}% Align table columns on decimal point
\usepackage{bm}% bold math

\def\lapproxeq{\lower .7ex\hbox{$\;\stackrel{\textstyle
<}{\sim}\;$}}
\def\gapproxeq{\lower .7ex\hbox{$\;\stackrel{\textstyle
>}{\sim}\;$}}

\begin{document}
% \begin{frontmatter}

\title{Practical and accurate calculations of Askaryan radiation.}

\author{Jaime Alvarez-Mu\~niz}
\affiliation{%
Depto. de F\'\i sica de Part\'\i culas
\& Instituto Galego de F\'\i sica de Altas Enerx\'\i as,
Universidade de Santiago de Compostela, 15782 Santiago
de Compostela, Spain
}%

\author{Andr\'es Romero-Wolf}
\affiliation{%
Jet Propulsion Laboratory, California Institute of Technology, 4800 Oak Grove Drive, 
Pasadena, California 91109, USA
}%

\author{Enrique Zas}
\affiliation{%
Depto. de F\'\i sica de Part\'\i culas
\& Instituto Galego de F\'\i sica de Altas Enerx\'\i as,
Universidad de Santiago de Compostela, 15782 Santiago
de Compostela, Spain
}%

\begin{abstract}
An in-depth characterization of coherent radio Cherenkov pulses from
particle showers in dense dielectric media, referred to as the
Askaryan effect, is presented. The time-domain calculation developed
in this article is based on a form factor to account for the lateral
dimensions of the shower. It is computationally efficient and able to reproduce the
results of detailed particle shower simulations with high fidelity
in most regions of practical interest including Fresnel
effects due to the longitudinal development of the shower. In
addition, an intuitive interpretation of the characteristics of the
Askaryan pulse is provided. We expect our approach to benefit the
analysis of radio pulses in experiments exploiting the radio technique.
\end{abstract}

%95.85.Bh Radio, microwave ( >1 mm)
%95.85.Ry Neutrino, muon, pion, and other elementary particles; cosmic rays
%29.40.-n Radiation detectors

% \keywords{high energy cosmic rays and neutrinos, high energy showers, Cherenkov radio emission}

\maketitle

%%%%%%%%%%%%%%%%%%%%%%%%%%%%%%%%%%%%%%%%%%
%\linenumbers

\section{Introduction}

In 1962 Askaryan~\cite{Askaryan62} proposed to detect Ultra-High Energy (UHE) cosmic rays and neutrinos observing the coherent radio pulse from the excess of electrons in a shower developing in a dense dielectric and non-absorptive medium. The scaling of the emitted power with the square of the particle energy, which has been experimentally confirmed in accelerators~\cite{Saltzberg_SLAC_sand,Miocinovic_SLAC_sand,Gorham_SLAC_salt,Gorham_SLAC_ice}, makes the technique very promising for the detection of UHE 
particles and has motivated a variety of past and present experiments~\cite{Parkes96,RICE03,GLUElimits,NuMoon,Kalyazin,ANITA_2009_limits,LUNASKA,RESUN}
along with some in the planning stages~\cite{ARA,ARIANNA}.
  
Key to the success of these initiatives is an accurate and computationally efficient calculation of the
radio emission properties due to the Askaryan effect in UHE showers.
The problem of computing the coherent radiation from particle showers can be approached in a variety of ways. 
Purely Monte Carlo methods have been developed to simulate the induced
showers in dense media. 
One can obtain the contribution to the radiation from every particle track in the shower from first principles
(Maxwell's equations), and add the contributions, which automatically takes coherent effects into account. 
This approach has been applied to calculate the Fourier components of the radiation (i.e. in the frequency-domain) 
\cite{ZHS91,ZHS92,alz97,alz98,alvz99,alvz00,razzaque01,almvz03,razzaque04,McKay_radio,alvz06,aljpz09,zhaires10}, 
and only recently to the calculation of the radiation as a function of time (i.e. in the time-domain) \cite{ARZ10}. 
Similar methods have also
been applied to the calculation of radio emission in atmospheric showers 
\cite{ReAires,Konstantinov,REAS3,SEIFAS,zhaires_nantes,zhaires_air11,SEIFAS} in which
the Askaryan effect is not the dominant mechanism and will not be
addressed in this paper. 
Monte Carlo methods have the advantage that the full complexity 
of shower phenomena is accounted for, the influence of shower-to-shower fluctuations can 
be addressed, and the dependence on the type of primary particle,
hadronic model, along with any other assumptions can be studied 
with high accuracy. However, purely Monte Carlo methods are typically very time-consuming, especially
at ultra-high energies and approximations are required \cite{aljpz09}.

Another numerical approach currently being developed is the application of finite difference in 
the time-domain (FDTD) techniques \cite{Chih10}. The idea is to discretize space-time and propagate the electric 
and magnetic fields by approximation of Maxwell's differential equations into difference equations \cite{FDTDbook}.  
FDTD techniques have the advantage that they can be easily adapted to computing the effects of dielectric 
boundaries and index of refraction gradients and can be linked to an accurate Monte Carlo simulation of 
showers in dense media. The FDTD technique is however rather computationally intensive \cite{Chih10}.

Analytical approaches have also been developed. In these methods 
the charge development in the shower is approximated as a current density vector \cite{buniy02}. Typically, 
parameterizations of the longitudinal and lateral profile of the showers are used to  
describe the main features of the space-time evolution of 
the charge distribution. In this approach one calculates the vector potential by 
integrating the Green's function to obtain the electric field. These integrals are in general 
difficult both numerically and analytically, but they can be greatly simplified by the use 
of approximations \cite{buniy02}. These methods are usually less time-consuming but 
also less accurate than purely Monte Carlo simulations. One limitation is that showers elongated due to the 
Landau-Pomeranchuk-Migdal (LPM) 
effect \cite{LPM,Stanev_LPM} cannot be parameterized easily due to large shower-to-shower 
fluctuations and the emitted radiation is known to depend strongly on the particular longitudinal 
profile of the shower \cite{alz97,alz98,alvz99}.
Analytical techniques have also been applied with different levels of sophistication
to the calculation of radio emission in atmospheric showers 
\cite{Huege_analytical,Scholten_MGMR,Montanet_radio,Ardouin_Coulomb}.

It is clearly important to be able to calculate coherent radiation due to the Askaryan as well
as other effects with a variety of techniques 
as each has its own set of advantages and disadvantages. 
Our goal in this work is to provide a calculation method  
that is both fast and able to reproduce all the essential characteristics resulting from 
detailed Monte Carlo shower simulations. Semi-analytical techniques are a very good option. The general idea behind this method is to obtain the charge distribution from detailed Monte Carlo simulations as the input for an analytical 
calculation of the radio pulse.
Other semi-analytical methods in dense media \cite{alz97,alz98,alvz99,ARZ10} and 
in the atmosphere \cite{Scholten_MGMR_sims} have been attempted in the past. 
In dense media, the frequency spectrum of the radio emission due to the Askaryan effect 
has been shown to be easily obtained from the Fourier-transform of the longitudinal profile of the shower~\cite{alvz99}. This technique reproduced the frequency spectrum as predicted in Monte Carlo simulations with
a high degree of accuracy, but only for angles away from the Cherenkov angle in the Fraunhofer approximation. 
Complementary, in the time-domain, it was shown that the electric field away from the Cherenkov angle and
in the far-field regime can be very accurately calculated from the time-derivative of the simulated longitudinal 
development of the excess charge~\cite{ARZ10} (see also \cite{Scholten_MGMR}).

This paper presents a semi-analytical calculation that is able to reproduce the electric field in the time-domain at all angles with respect to the shower axis in both the far-field (Fraunhofer) and ``near-field" \footnote{By ``near-field" we mean a region
in which the Fraunhofer approximation is not valid because of the longitudinal dimensions of the shower and not 
the region where the Coulomb field associated with the charge excess cannot be considered negligible.} 
regions of the shower when compared to a full Monte Carlo simulation such as 
the well-known Zas-Halzen-Stanev (ZHS) code \cite{ZHS92,ARZ10,ZHSinprep}. 
The technique is computationally 
efficient since it only requires the convolution of the longitudinal charge excess profile with a parameterized {\it form factor} 
to fully reproduce the coherent radiation effects from particle showers in a homogeneous dielectric medium. 
Once the longitudinal shower profile is obtained, the electric field can be 
calculated with a simple numerical integral. 
It is worth remarking here that the longitudinal development of extremely energetic showers can be obtained quickly and with high precision, using hybrid simulation techniques which consist on following only the highest energy particles in the shower while accounting for the lowest energy particles with parameterizations. 
In particular, the complexity of the longitudinal profile of showers affected by the LPM effect can be very well reproduced with hybrid techniques \cite{alz97,alz98,alvz99} (for an example see Fig.~$5$ in \cite{aljpz09}).

The semi-analytical method described in this work is well suited to obtain the time-domain radio emission due to electromagnetic showers, for all observation angles both in the far-field and the near-field regions of the shower. 
The approach can be used in practically all experimental situations of interest 
since it only begins to show significant discrepancies when the observer
is at distances comparable to the lateral dimensions of the shower
($\lesssim$~1~m in ice). Since the typical distance between antennas in experiments
such as the Askaryan Radio Array (ARA) \cite{ARA} is $\sim 10 - 100$ m,
we expect the results to be accurate enough in most practical situations. 

We expect our results to benefit experiments exploiting the radio
technique. They can be used in detector simulations to test the
efficiency for pulses observed from various directions. In particular,
with our approach one can test the ability to detect the craggy pulses resulting from the
LPM effect. The calculation can also be implemented in the data
analysis of experiments by using likelihood functions aimed at the
reconstruction of the longitudinal charge excess profile from a detected pulse. 
If the longitudinal distribution is consistent with the elongation and
multiple-peaked structure due to the LPM effect, this can be used for neutrino flavor 
identification since it is only expected in UHE showers due to electron neutrinos.

%%%%%%%%%%%%%%%%%%%%%%%%%%%%%%%%%%%%%%%%%%%%%%%%%%%%%%%%%%%%%%%%%%%%%%%%%%%%%%

\section{Modeling Askaryan radiation}

The case we are interested in this paper is the radiation due to the charge excess of a shower in a linear dielectric medium
such as ice, salt, or silica sand.
We use SI units all throughout this work.
The Green's function solutions to Maxwell's equations provide the potentials $\Phi$ and $\mathbf{A}$ given a charge 
distribution $\rho$ 
with current density vector $\mathbf{J}=\rho\mathbf{v}$. Assuming a dielectric constant $\epsilon$ and magnetic 
constant $\mu$ the solutions in the Coulomb gauge ($\nabla\cdot\mathbf{A}=0$) can be written as 
\begin{equation}
\Phi(\mathbf{x},t)=\frac{1}{4\pi\epsilon}\int_{-\infty}^{\infty}  
\frac{\rho(\mathbf{x}',t)}{|\mathbf{x}-\mathbf{x}'|} d^3\mathbf{x}'
\end{equation} 
\begin{equation}
\mathbf{A}(\mathbf{x},t)=\frac{\mu}{4\pi}\int_{-\infty}^{\infty}  
\frac{\mathbf{J}_{\perp}(\mathbf{x}',t')}{|\mathbf{x}-\mathbf{x}'|}
\delta\left(\sqrt{\mu\epsilon}|\mathbf{x}-\mathbf{x}'|-(t-t')\right) d^3\mathbf{x}' dt'
\label{VectorPotential}
\end{equation} 
where the delta function gives the observer's time $t$ delayed with respect to the source time $t'$ 
by the time it takes light to reach the observation point $\mathbf{x}$ from 
the source position at $\mathbf{x}'$. The transverse current is given 
by  $\mathbf{J}_{\perp}=-\mathbf{\hat{u}}\times\-(\mathbf{\hat{u}}\times\mathbf{J})$ 
where $\mathbf{\hat{u}}=(\mathbf{x}-\mathbf{x}')/|\mathbf{x}-\mathbf{x}'|$ is the unit vector pointing from the source to the observer.
The non-trivial proof that the $\mathbf{J}_{\perp}$ above is the only component
relevant to the radiation part of the field is given in~\cite{Jackson_AmJPhys}.

Radiation calculations are typically performed in the Lorentz gauge 
($\nabla\cdot\mathbf{A} + n^2~c^{-2}\partial\Phi/\partial t=0$)
~\cite{Jackson,buniy02,razzaque04,Scholten_MGMR}, 
with $n$ the refractive index of the medium.
However, our primary interest is to derive an 
approach that can be easily applied to numerical radiation calculations. In the Coulomb gauge, the scalar potential 
only describes near-field terms which can be ignored for our purposes. This simplifies the computation of the radiative electric field $\mathbf{E}=-\nabla\Phi-\partial\mathbf{A}/\partial t$ to a simple time derivative 
$\mathbf{E}=-\partial\mathbf{A}/\partial t$. Thus, all that is needed is a computation of the vector potential 
as a function of time at the position of interest.

\begin{figure}[tb]
\begin{center}
\includegraphics[width=\linewidth]{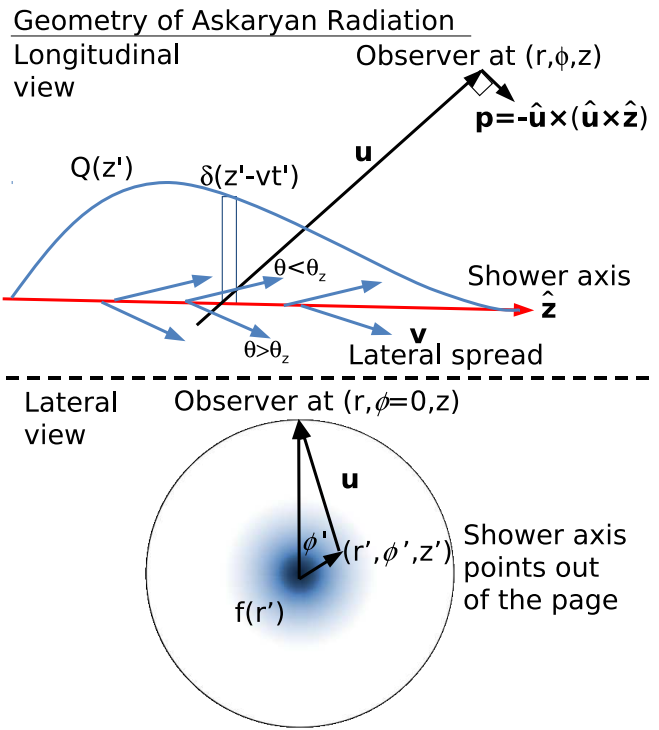}
\caption{Geometry of a high energy particle shower. The top figure shows a side view ($r-z$ plane) of the charge excess profile. The shower front propagates at velocity $v$ and is modeled as a thin pancake $\delta(z'-vt')$. The charge evolution of the shower front traces out the longitudinal profile $Q(z')$ shown here as an asymmetric distribution.  The lateral spread of the shower due to Coulomb scattering, shown in (blue) arrows, is modeled by a velocity vector $\mathbf{v}=\mathbf{v}(r',\phi')$ that is mostly directed along the shower axis with small radial components. The scatter results in observation angles that differ from the nominal angle $\theta_z$ relative to the shower axis and will lead to an asymmetric pulse. The radiative portion of the electric field lies along the vector $\mathbf{p}$ which is the orthogonal projection of $\mathbf{\hat{z}}$ along the direction $\mathbf{\hat{u}}$ from the source to the observer. The bottom panel shows a frontal view ($r-\phi$ plane) of the shower. The lateral charge excess distribution $f(r')$ is represented in (blue) shading.}
\label{fig:geom}
\end{center}
\end{figure}

The radiation of a particle shower in a dense medium is obtained by treating it as a current density $\mathbf{J}$ 
with its main features depicted in Fig.\ref{fig:geom}. 
The evolution in space-time of the excess charge in a shower can be modeled as a pancake $\delta(z'-vt')$ traveling 
with velocity $\mathbf{v}$ along the $z$--axis. The net charge profile of the shower $Q(z')$ rises and falls along 
the shower direction $z'$ and spreads laterally in $x'$ and $y'$. The velocity vector $\mathbf{v}$ is primarily 
directed in the shower axis direction $\mathbf{\hat z}$ but may have a small lateral component and a lateral 
dependence due to scattering of particles in the shower. This may seem like an unnecessary complication but 
the small scatter will lead to an observable asymmetry of the Askaryan pulse in the time-domain. The speed $v$ 
is assumed to be close to the speed of light for particle showers of interest.
The associated current density vector can be modeled as a cylindrically symmetric function
\begin{equation}
\mathbf{J}(\mathbf{x}',t')=\mathbf{v}(r',\phi',z')f(r',z')Q(z')\delta(z'-vt')
\end{equation}
where $r'=\sqrt{x'^2+y'^2}$ is the cylindrical radius. The function $f(r',z')$ represents the lateral charge 
distribution in a plane 
transverse to $z'$ depicted in the bottom of Fig.~\ref{fig:geom}. This model of the current density 
is similar to the ones used 
in~\cite{buniy02,razzaque04} except that $\mathbf{v}$ is allowed to have first order radial components.

The vector potential is then given by
\begin{equation}
\begin{split}
&
\mathbf{A}(\mathbf{x},t)=\frac{\mu}{4\pi}
\int^{\infty}_{-\infty} dt'
\int^{\infty}_{-\infty} dz' Q(z') \delta(z'-vt')
\\
& 
\int^{\infty}_{0} dr' r' \int^{2\pi}_{0} d\phi'  f(r',z')  \mathbf{{v}}_{\perp}(r',\phi',z') 
\\
&
\frac{\delta\left(n|\mathbf{x}-\mathbf{x}'|/c-(t-t')\right)}{|\mathbf{x}-\mathbf{x}'|} 
\end{split}
\label{eq:vp}
\end{equation} 
where $\phi'$ is the azimuthal angle in cylindrical coordinates and
$\mathbf{v}_{\perp}(r',\phi', z')$ generally depends on $r'$, $\phi'$, and $z'$.

In its full glory Eq.~(\ref{eq:vp}) seems rather intractable. 
The lateral distribution $f(r',z')$ is the most difficult part to model. 
It is for this reason that the semi-analytical models developed in \cite{alvz99} and \cite{ARZ10} ignored the lateral development of the shower, and hence were not able to accurately describe Askaryan radiation near the Cherenkov angle where the lateral distribution is known to determine the degree of coherence of the emitted radiation \cite{ZHS92}.
Modeling the radiation due to the lateral distribution of the shower can be attempted using 
the standard NKG function \cite{Greisen60}. However, this parameterization has 
singularities that make the integrals particularly difficult to solve and interpret,
and it assumes constant particle velocities parallel to the shower axis which is 
not sufficiently accurate as will be shown further below. 

In the following we will show that the 
radiation due to the lateral distribution of the shower can be parameterized, and 
this parameterization can be used to predict  
the emitted radiation matching that obtained from the full ZHS simulation \cite{ZHS92,ARZ10} 
with great accuracy. 
Moreover, the lateral distribution of the shower is mainly due to 
low energy processes such as Coulomb scattering, 
and as result the shape of the parameterized radiation is independent of shower energy 
in the energy range of interest as will be shown below. 
On the other hand, the longitudinal distribution $Q(z')$ 
can change dramatically depending on the energy 
of the shower due to the LPM effect, and needs to be obtained in a Monte Carlo simulations such as ZHS.
We will also show that our calculation of Askaryan radiation works in both the near and far-field approximations,
as long as the lateral coordinates are treated in the far-field, i.e. the observation of the shower occurs
at a distance larger than its lateral dimensions ($\sim$~1~m in ice).

%%%%%%%%%%%%%%%%%%%%%%%%%%%%%%%%%%%%%%%%%%%%%%%%%%%%%%%%%%%%%%
\subsection{The vector potential at the Cherenkov angle}

Let us first consider the Fraunhofer approximation for radiation
emitted by a shower. This implies expanding $|\mathbf{x}-\mathbf{x}'|$ in
Eq.~(\ref{VectorPotential}) as  
\begin{equation}
|\mathbf{x}-\mathbf{x}'|\approx R-\mathbf{\frac{x}{\vert x\vert}}\cdot\mathbf{x}'
\end{equation}
where $R=\mathbf{\vert x \vert}$. For an observer looking in the direction 
$\mathbf{\hat{u}}=\mathbf{x/\vert x\vert}=(\sin\theta\cos\phi,\sin\theta\sin\phi,\cos\theta)$
in spherical coordinates, and assuming without loss of generality that $\phi=0$, the 
above expansion can be written as
\begin{equation}
|\mathbf{x}-\mathbf{x}'|\approx R-z'\cos\theta-r'\sin\theta\cos\phi'
\label{eq:approx_fra}
\end{equation}
Approximating the denominator of the vector potential in Eq.~(\ref{eq:vp})
by $|\mathbf{x}-\mathbf{x}'|\approx R$ but keeping the approximation 
in Eq.~(\ref{eq:approx_fra}) in the argument of the $\delta$-function we get, 
\begin{equation}
\begin{split}
&
\mathbf{A}(\theta,t)=\frac{\mu}{4\pi R} 
\int^{\infty}_{-\infty}dt'
\int^{\infty}_{-\infty}dz' Q(z')\delta(z'-vt')
\\
&
\int^{\infty}_{0} dr' r' \int^{2\pi}_{0} d\phi'  f(r',z') \mathbf{{v}}_{\perp}(r',\phi',z') \\
&
\delta\left(\frac{n[R-z'\cos\theta-r'\sin\theta\cos\phi']}{c}-(t-t')\right)
\end{split}
\end{equation} 
Integrating over the source time $t'$ results in
\begin{equation}
\begin{split}
&
\mathbf{A}(\theta,t)=
\frac{\mu}{4\pi R}  
\int^{\infty}_{-\infty}dz'Q(z')
\int^{\infty}_{0} dr' r' f(r',z')
\\
& 
\int^{2\pi}_{0} d\phi'
 \frac{\mathbf{v}_{\perp}(r',\phi',z')}{v} 
\\
&
\delta\left(z'\left[\frac{1}{v}-\frac{n\cos\theta}{c}\right]-\frac{nr'\sin\theta\cos\phi'}{c}+\frac{nR}{c}-t\right).
\end{split}
\label{eq:vp_angle}
\end{equation} 
where $\mathbf{v}_{\perp}(r',\phi',z') = -\mathbf{\hat{u}}\times\mathbf{\hat{u}}\times\mathbf{v}(r',\phi',z')$ 
is the transverse projection of the unit velocity vector.

In our model we make the assumption that the shape of the lateral density
and the particle velocity depend only very weakly on $z'$,
i.e. $f(r',z')\simeq f(r')$ and  $\mathbf{v}(r',\phi',z') \simeq
\mathbf{v}(r',\phi')$. 
With these approximations Eq.~(\ref{eq:vp_angle}) can be written as:
\begin{equation}
\begin{split}
&
\mathbf{A}(\theta,t)=\frac{\mu}{4\pi R}~\sin\theta~
\int^{\infty}_{-\infty} dz'Q(z')
\\
&
\mathbf{F}\left(t-\frac{nR}{c} - z'\left[\frac{1}{v}-\frac{n\cos\theta}{c}\right]\right)~
\end{split}
\label{AvsLambda_angle} 
\end{equation} 
where we have defined the function $\mathbf{F}$ as:
\begin{equation}
\begin{split}
&
\mathbf{F}\left(t-\frac{nR}{c} - z'\left[\frac{1}{v}-\frac{n\cos\theta}{c}\right]\right) = 
\\
&
\frac{1}{\sin\theta}~
\int^{\infty}_{0} dr' r' \int^{2\pi}_{0} d\phi' f(r')
\frac{\mathbf{v}_{\perp}(r',\phi')}{v} 
\\
& 
\delta\left(z'\left[\frac{1}{v}-\frac{n\cos\theta}{c}\right]-\frac{nr'\sin\theta\cos\phi'}{c}+\frac{nR}{c}-t\right)
\end{split}
\label{form_factor_angle}
\end{equation}
Note that we have explicitly included a factor $\sin\theta$ in Eq.~(\ref{AvsLambda_angle}) for convenience, 
anticipating that the radiation is mainly polarized in the direction transverse to the observer's direction.

The vector function $\mathbf{F}$ contains the radial $r'$ and 
azimuthal $\phi'$ integrals and can be considered  
as an effective form factor that accounts for the lateral distribution of the charged current density, 
quite analogous to that obtained in the frequency domain in \cite{buniy02}.

At the Cherenkov angle we have $1/v-n\cos\theta_C/c=0$ and Eq.~(\ref{AvsLambda_angle}) results in:
%%%%%%%%%%%%%%%%%%%%%%%%%%%%%%%%%%%%%%%%%
\begin{figure}[tb]
\begin{center}
\includegraphics[width=\linewidth]{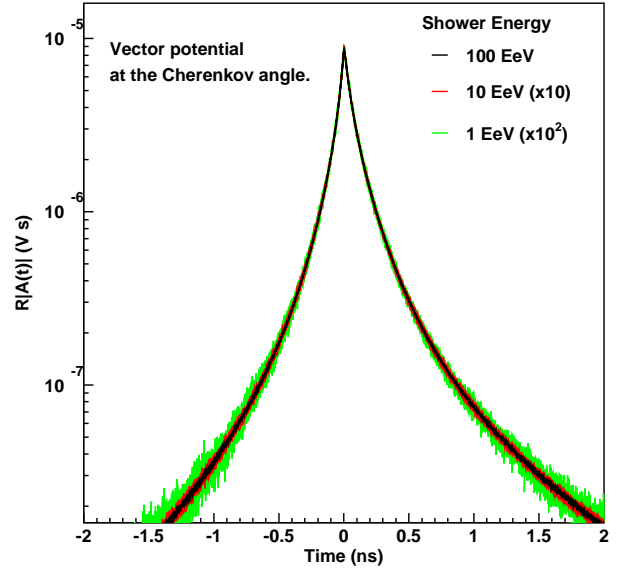}
\caption{{ The vector potential from electromagnetic showers 
in homogeneous ice (density $\rho=0.924~{\rm g~cm^{-3}}$ and refractive index $n=1.78$, $\theta_C\sim55.8^\circ$)
observed at the Cherenkov angle for various energies from the ZHS simulation. 
The functional behavior is identical except for an overall scaling factor that is directly proportional to the shower energy. The width is determined by the lateral distribution of the shower while the asymmetry is mainly due to the radial spread of particle tracks; both are the result of the Coulomb scattering in the medium.}}
\label{fig:vp_cer}
\end{center}
\end{figure}
%%%%%%%%%%%%%%%%%%%%%%%%%%%%%%%%%%%%%%%%%%
%
\begin{equation}
\mathbf{A}(\theta_C,t)=\frac{\mu}{4\pi R}~\sin\theta_C~\mathbf{F}\left(t-\frac{nR}{c}\right)~
\int^{\infty}_{-\infty} dz'Q(z')
\label{AvsLambda} 
\end{equation} 
where $\mathbf{F}$ at the Cherenkov angle is obtained from Eq.~(\ref{form_factor_angle}): 
\begin{equation}
\begin{split}
&
\mathbf{F}\left(t-\frac{nR}{c}\right) = 
\frac{1}{\sin\theta_C}~
\int^{\infty}_{0} dr' r' f(r') \int^{2\pi}_{0} d\phi'  
\frac{\mathbf{v}_{\perp}(r',\phi')}{v}
\\
&
\delta\left(\frac{nR}{c}-t-\frac{nr'\sin\theta_C\cos\phi'}{c}\right).
\end{split}
\label{eq:def_lambda}
\end{equation}
%
%in units of $s^{-1}$.

Using symmetry arguments we can now project the form factor  
$\mathbf{F}$ in only two orthogonal directions using unit vectors 
$\mathbf{\hat{u}}$ along the observation direction and
in the direction of $\mathbf{p}=-\mathbf{\hat{u}} \times (\mathbf{\hat{u}} \times\mathbf{\hat{z}})$: 
\begin{equation}
\mathbf{F}=F_p ~\mathbf{\hat{p}} + F_u \mathbf{\hat{u}} 
\label{eq:Fcomponents}
\end{equation} 
The direction of $\mathbf{\hat{p}}$ has been chosen perpendicular to
the observation direction and lying on the plane defined by 
$\mathbf{\hat{z}}$ and $\mathbf{\hat{u}}$ as shown in
Fig.~\ref{fig:geom}\footnote{By use of vector identities the direction
of $\mathbf{\hat{p}}$ can be shown to be that of
$\mathbf{\hat{z}}-(\mathbf{\hat{z}}\cdot\mathbf{\hat{u}})\mathbf{\hat{u}}$
which is more apparent.}. This has been done anticipating the expected polarization
of the radiation mainly in the direction of $\mathbf{\hat{p}}$ in order to
make the orthogonal component along the direction of observation $F_u$ negligible. 

The form factor $\mathbf{F}$ is a medium dependent 
function that accounts in an effective way for the radial and azimuthal 
interference effects due to the lateral structure of the shower
including possible directional variations in the velocity vector
$\mathbf{v}$ of the shower particles. 
In principle $\mathbf{F}$ can be obtained from analytical
solutions of the cascade
equations, but this is rather involved. Alternatively one could use
standard parameterizations of the lateral distribution function of high energy
showers such as the NKG ~\cite{Greisen60}. However, as stated before, this does not
account for the radial components of the velocity and gives results
symmetric in time which are qualitatively different from results
obtained in detailed simulations such as that shown in Fig.~\ref{fig:vp_cer}. 
The trick is to extract $\mathbf{F}$ from simulations that effectively
account for the radial component of particle velocities and the
lateral distribution of the excess charge which are responsible for
the time asymmetry characteristic of simulations. 
The basic idea behind this article is that the form factor is obtained
from the vector potential at the Cherenkov angle and, as we will show later,
the emission at other angles is easily related to that at the Cherenkov angle.

The vector potential at the Cherenkov angle in the time domain is
calculated with a detailed shower simulation and parameterized for
practical purposes. The form factor 
$\mathbf{F}$ can be obtained directly equating Eq.~(\ref{AvsLambda}) to the
vector potential as obtained in the simulation. As anticipated the $F_u$ 
component is typically below $1~\%$ of $F_p$ and it can be neglected, and 
$F_p$ can be simply obtained from the following equation:
\begin{equation}
\mathbf{A}(\theta_C,t)=\frac{\mu}{4\pi R }~
\sin\theta_C~F_p\left(t-\frac{nR}{c}\right)~LQ_{\rm tot}~\mathbf{\hat{p}}
\label{eq:vp_cher}
\end{equation}
where $LQ_{\rm tot}=\int dz'Q(z')$ has been referred
to as the excess projected track-length \cite{ZHS92}. 
Taking the absolute value of Eq.~(\ref{eq:vp_cher}) the functional form of $F_p$ is given by:
\begin{equation}
F_p\left(t-\frac{nR}{c}\right) = \frac{4\pi}{\mu}~\frac{RA(\theta_C,t)}{LQ_{\rm tot}}~
\frac{1}{\sin\theta_C}
\label{eq:lambda}
\end{equation}
where $A(\theta_C,t)=|\mathbf{A}(\theta_C,t)|$. $F_p$ represents the average vector potential at the Cherenkov angle per unit excess track length - given by $LQ_{\rm tot}$ -
scaled with the factor $4\pi R/\mu$. 

Detailed simulations of electromagnetic showers performed with the ZHS Monte Carlo code in ice produce a consistent 
time-dependent vector potential at all energies of interest as shown in Fig.~\ref{fig:vp_cer}. 
The results for homogeneous ice can be parameterized by:
\begin{equation}
\begin{split}
&
RA(\theta_C,t) = -4.5\times 10^{-14}~[\mbox{V s}] ~ \frac{E}{[\mbox{TeV}]}
\\
&
\left\{ 
\begin{array}{l l}
\exp\left(-\frac{|t|}{0.057}\right)+(1+2.87|t|)^{-3} & \quad \mbox{if $t>0$}\\ \\
\exp\left(-\frac{|t|}{0.030}\right)+(1+3.05|t|)^{-3.5} & \quad \mbox{if $t<0$}\\ 
\end{array} 
\right. 
\end{split}
\label{eq:vp_fit}
\end{equation}
where $E$ is the energy of the shower in TeV and $t$ is the observer time in ns. The result is accurate to within 5\%.
The shape of $A(\theta_C,t)$ depends very weakly on shower energy, while the 
normalization is proportional to the energy as becomes evident in Fig.~\ref{fig:vp_cer}.
Note also that at higher energies the fluctuations are reduced because the number
of particle tracks increases almost linearly with shower energy.  

%%%%%%%%%%%%%%%%%%%%%%%%%%%%%%%%%%%%%%%%%%%%%%%%%%%%%%%%%%%%%%%%%%%%%%%%%
\subsection{The radiation in the far-field}

Given the time domain parametrization of the radiation at the Cherenkov angle in Eq.~(\ref{eq:vp_fit}) 
we will first obtain the pulse as seen by an observer in
the far-field at any observation angle. 
The integral in Eq.~(\ref{AvsLambda_angle}) can be written as
\begin{equation}
\begin{split}
&
\mathbf{A}(\theta,t)=\frac{\mu}{4\pi R}~\sin\theta~\mathbf{\hat{p}} 
\int^{\infty}_{-\infty}dz'Q(z') 
\\
&
F_p\left(t-\frac{nR}{c}-z'\left[\frac{1}{v}-\frac{n\cos\theta}{c}\right]\right).
\end{split} 
\label{eq:model_Fraun}
\end{equation} 
%
%The argument of the form factor $F_p$ now includes an
%extra term, $z'(1/v-n \cos \theta/c)$, which vanishes in the Cherenkov angle.
This is a straightforward numerical integration with $F_p$ given
by Eqs.~(\ref{eq:lambda}) and (\ref{eq:vp_fit}) and the longitudinal distribution 
of the excess charge $Q(z')$ which can be obtained in a shower simulation or
from a standard parameterization of the depth development if the LPM
effect is absent. The corresponding radiative electric field
is obtained by simply taking the derivative of the vector potential with respect to time $\mathbf{E}=-\partial{\mathbf A}/\partial t$.

Eq.~(\ref{eq:model_Fraun}) tells us that the vector potential in the far-field region of the shower 
can be obtained as a convolution of the form factor $F_p$,  
which parameterizes the emission from the lateral distribution of the shower, with the longitudinal profile 
of the excess charge. The form factor $F_p$ is a function that has to be evaluated at 
the time $t$ at which the observer in 
the far field sees the portion of the shower corresponding to the
depth $z'$. That time is given by $t = nR/c + z'/v - z'n\cos\theta/c$. 
We have made the only assumption 
that the shape of $F_p$ depends weakly on the stage of longitudinal evolution of
the shower. At the Cherenkov angle, 
the far-field observer sees the whole longitudinal development of the shower at once 
i.e. $z'/v = z'n\cos\theta_C/c$ in which case Eq.~(\ref{eq:model_Fraun}) reduces to the vector potential at the Cherenkov angle given by Eq.~(\ref{eq:vp_cher}).

%%%%%%%%%%%%%%%%%%%%%%%%%%%%%%%%%%%%%%%%%
\begin{figure}[tb]
\begin{center}
\includegraphics[width=\linewidth]{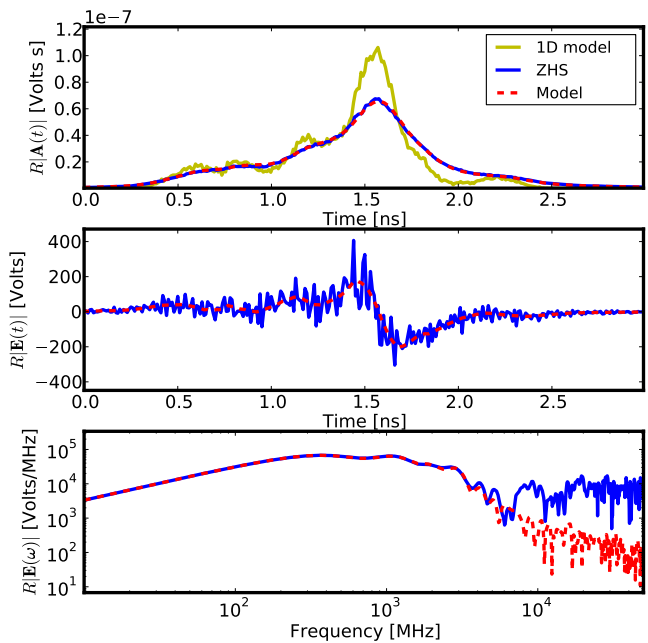}
\caption{Results of the ZHS Monte Carlo compared to our calculation Eq.~(\ref{eq:model_Fraun}) for the radiation due to a $3\times10^{18}$ eV electromagnetic shower observed at $\theta=\theta_C-0.3^{\circ}$ in the far-field in ice.
The method is applied to the charge excess longitudinal profile obtained from the same simulation. Top panel: Vector potential in the time-domain with 10~ps 
time sampling (corresponding to a sampling frequency of 100 GHz) as obtained in ZHS simulations. This is compared 
to the calculation presented in this work. The longitudinal charge profile is shown as the 1D model presented in \cite{ARZ10} where the depth and charge are linearly rescaled to give the observer time and vector potential. 
Middle panel: Electric field in the time-domain comparing our method to ZHS results. 
Bottom panel: The electric field amplitude spectrum obtained from the ZHS
simulation and the Fourier transform amplitudes of the electric field
obtained from the calculation presented in this work. Note that the
discrepancy between the time-domain electric field of the ZHS
simulations and our results are due to the incoherent radiation at high frequencies.}
\label{fig:LPM}
\end{center}
\end{figure}
%%%%%%%%%%%%%%%%%%%%%%%%%%%%%%%%%%%%%%%%%%

In Fig.~\ref{fig:LPM} we show an example of the vector potential and electric field 
in the time-domain due to an electromagnetic shower with energy
$E=100$ EeV from the ZHS simulation. The fields are observed in the
Fraunhofer region at an angle $\theta=\theta_C - 0.3^\circ$ and they
are compared to our results obtained with Eq.~(\ref{eq:model_Fraun})
using the longitudinal distribution $Q(z')$ from the same simulation.
The agreement between the vector potential obtained directly in the
Monte Carlo simulation 
and the prediction of Eq.~(\ref{eq:model_Fraun}) lies within a few percent difference in the region relevant to the pulse (top panel of Fig.~\ref{fig:LPM}). 
The difference between this calculation and the ZHS electric field in the time domain 
is greater (middle panel of Fig.~\ref{fig:LPM}), but as shown in the
bottom of Fig.~\ref{fig:LPM} this is due mostly to the incoherent emission 
of the shower at high frequencies. 
The Fourier-transformed amplitudes of the time-domain electric field obtained
in our approach are based on smooth parameterizations, 
while the frequency spectrum obtained directly in the 
ZHS simulation includes incoherence effects coming from the fine 
structure of the shower at the individual particle level. 

Note that when $f(r',z')=\delta(r')/r'$, i.e. if we neglect the lateral distribution 
of the shower, Eq.~(\ref{eq:model_Fraun}) reduces to the 1-dimensional model in \cite{ARZ10}
which fails at describing the features of the radio emission for angles close
to the Cherenkov angle. To illustrate this, we show in the top panel of Fig.~\ref{fig:LPM}
the vector potential obtained in the one~dimensional~(1D) model in \cite{ARZ10}, 
which is a linear rescaling of the longitudinal charge excess profile,
showing a clear disagreement with the results of the full ZHS simulation as expected.

%%%%%%%%%%%%%%%%%%%%%%%%%%%%%%%%%%%%%%%%%%%%%%%%%%%%%%%%%%%%%%%%%%%%%%%%%
\subsection{Askaryan pulses in the ``near-field"} 
\label{Near-field}

We can now generalize Eq.~(\ref{eq:model_Fraun}) for an observer in the ``near-field" region of the shower. 
In this case it is more natural to work in cylindrical coordinates and place the observer
at $(r\cos\phi,r\sin\phi,z)$. Without loss of generality we can again assume the observer is at $\phi=0$ giving
\begin{equation}
|\mathbf{x}-\mathbf{x}'|= \sqrt{r^2+r'^2-2rr'\cos\phi'+(z-z')^2}.
\label{eq:Fresnel_approx}
\end{equation}
In dense media, the lateral distribution is in the scale of centimeters, which means that for all 
practical purposes the observer is at any given instant in the
far-field region with respect to the lateral distribution. 
The idea is to solve the vector potential in Eq.~(\ref{eq:vp}) using the Fraunhofer
approximation to account for the lateral distribution at any given
time $t'$. 
We expand Eq.~(\ref{eq:Fresnel_approx}) to first order in $r'$ giving
\begin{equation}
|\mathbf{x}-\mathbf{x}'|= \sqrt{r^2+(z-z')^2}-r'\sin\theta(z')\cos\phi'.
\label{eq:R_Fresnel}
\end{equation}
where $\sin\theta(z') = r/\sqrt{r^2+(z-z')^2}$, but we take into account
that the distance in the denominator 
of the vector potential depends on the time $t'$ or equivalently
on the position $z'$ in the shower as $\sqrt{r^2+(z-z')^2}$.
This is in contrast to the case of the far field calculation in which the distance in the 
denominator of the vector potential is constant and equal to $R$. 

We proceed as in the case of the far-field calculation assuming that 
$f(r',z')\simeq f(r')$ and $\mathbf{v}(r',\phi',z') \simeq \mathbf{v}(r',\phi')$. 
After integrating over $t'$ the vector potential in Eq.~(\ref{eq:vp}) can be written as, 
\begin{equation}
\begin{split}
&
\mathbf{A}(r,z,t)= \frac{\mu}{4\pi}
\int_{-\infty}^{\infty} dz'\frac{Q(z')}{\sqrt{r^2+(z-z')^2}} 
\\
&
\int_{-\infty}^{\infty} dr' r' \int_{0}^{2\pi} d\phi' f(r') ~\frac{\mathbf{v}_{\perp}(r',\phi',z')} {v} 
\\
& 
\delta\left(\frac{z'}{v} + \frac{n\sqrt{r^2+(z-z')^2}-nr'\sin\theta(z') \cos\phi'}{c}-t\right)
\label{eq:vp_Fresnel}
\end{split}
\end{equation}
where the transverse projection of the velocity vector
$\mathbf{v}_{\perp}(r',\phi',z')
=-\mathbf{\hat{u}}(z')\times[\mathbf{\hat{u}}(z')\times\mathbf{v}(r',\phi')]$
now introduces a new dependence on the longitudinal source coordinate $z'$ 
due to the fact that in the near-field 
$\mathbf{\hat{u}}(z')=(\mathbf{r}+(z-z')\mathbf{\hat{z}})/\sqrt{r^2+(z-z')^2}$ depends on $z'$. 
If we define the form factor containing the radial $r'$ and azimuthal
$\phi'$ integrals as in subsection II.A we obtain a similar 
expression:
\begin{equation}
\begin{split}
&
{\mathbf F}\left(t-\frac{z'}{v}-\frac{n\sqrt{r^2+(z-z')^2}}{c}\right) =
%   &
\\
& 
\int_{-\infty}^{\infty} dr'r' \int_{0}^{2\pi} d\phi' ~ f(r') ~\frac{\mathbf{v}_{\perp}(r',\phi',z')}{v}
\\
& 
\delta\left(\frac{z'}{v} + \frac{n\sqrt{r^2+(z-z')^2}}{c}-t-\frac{nr'\sin\theta(z')\cos\phi'}{c}\right)
\end{split}
\label{eq:lambda_Fresnel}
\end{equation}  
%

%%%%%%%%%%%%%%%%%%%%%%%%%%%%%%%%%%%%%%%%%
\begin{figure}[tb]
\begin{center}
\includegraphics[width=\linewidth]{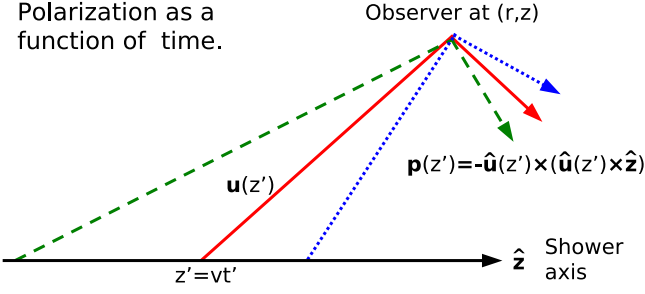}
\caption{ 
Sketch illustrating the polarization dependence with time. 
As the shower evolves the observer sees the radiation contributions 
at different angles corresponding to different source positions $z'$. This means that the observer will see the polarization vector changing as a function of time $t$.}
\label{fig:polarization}
\end{center}
\end{figure}
%%%%%%%%%%%%%%%%%%%%%%%%%%%%%%%%%%%%%%%%%%

Note that the form factor defined in Eq.~(\ref{eq:lambda_Fresnel}) has
the same functional form as that defined in Eq.~(\ref{eq:lambda}) and
they only differ in the argument of the delta function. 
The form factor $\mathbf{F}$ as obtained in the far field can
be applied to the near field (in relation to the longitudinal
development of the shower) simply modifying its argument.

Neglecting again the component of $\mathbf{F}$ parallel to $\mathbf{\hat{u}}$ the vector
potential in the near-field can be written as:
\begin{equation}
\begin{split}
&
\mathbf{A}(r,z,t)=  
\frac{\mu}{4\pi}
\int_{-\infty}^{\infty}
dz'\frac{Q(z')}{\sqrt{r^2+(z-z')^2}}~\mathbf{p}(z') 
\\
&
F_p\left(t-\frac{z'}{v}-\frac{n\sqrt{r^2+(z-z')^2}}{c}\right)
\label{eq:model_Fres}
\end{split}
\end{equation}

Note that a new $z'$ dependence is introduced through the polarization vector
$\mathbf{p}(z')=-\mathbf{\hat{u}}(z')\times(\mathbf{\hat{u}}(z')\times\mathbf{\hat{z}})$. 
In Fig.~\ref{fig:polarization} an example is shown of the $z'$ 
dependence of $\mathbf{p}(z')$. 
Also note that the $\sin\theta$ term in Eq.~(\ref{eq:model_Fraun}) has been absorbed 
in Eq.~(\ref{eq:model_Fres}) through $\mathbf{p}=\sin\theta~\mathbf{\hat p}$

In the near field the explicit $z'$ dependence is necessary because
the longitudinal profile of the shower 
is observed at different angles. This means that parts of the shower observed at different 
depths will have differing polarization vectors. 
This modification accounts exactly 
for the interference between different $z'$ points along the
shower development. This result has also been determined from a one
dimensional current density model in~\cite{Chen_ARENA_2010}.

Eq.~(\ref{eq:model_Fres}) tells us that the vector potential in the near-field region of the shower 
can be obtained as a convolution of the form factor $F_p$ - that
parameterizes the interference effects
due to the lateral distribution of the shower - and the longitudinal profile of the excess charge. 
The form factor function $F_p$ has to be evaluated at the time $t$ at
which an observer in the near field
sees the portion of the shower corresponding to the depth $z'$. That
time is clearly given by 
$t = z'/v + n\sqrt{r^2+(z-z')^2}/c$. The difference between this
expression and that obtained in the far field 
is that $F_p$ is always evaluated at the time $t$ an observer sees the 
position $z'$ in the shower which is different for the far- and
near-field regions. 

%We refer to this computation as the convolution model.

%%%%%%%%%%%%%%%%%%%%%%%%%%%%%%%%%%%%%%%%%
\begin{figure}[tb]
\begin{center}
\includegraphics[width=\linewidth]{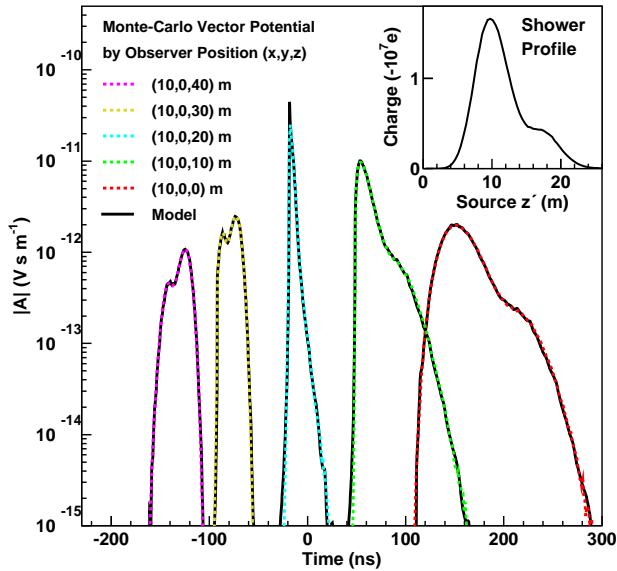}
\caption{ 
The Fresnel region vector potential for an electron induced $100$~PeV
shower in ice from the ZHS Monte Carlo compared to the results of Eq.~(\ref{eq:model_Fres}). 
The longitudinal profile of the excess charge is convolved with the
form factor $F_p$. The observers are located at $(x,y,z)$ where $z$ lies along the shower axis and $y=0$. The vector potentials have been arbitrarily shifted in time by multiples of 50~ns for clarity. The spiked vector potential in the middle (blue) corresponds to an observation where the peak of the longitudinal shower profile is near the Cherenkov angle. The vector potentials to the right of it (green and red) are due to observations above the Cherenkov angle while the vector potentials to the left (magenta and yellow) are due to observations below the Cherenkov angle. Note the inverted order of the primary and secondary peaks.}
\label{fig:Fresnel}
\end{center}
\end{figure}
%%%%%%%%%%%%%%%%%%%%%%%%%%%%%%%%%%%%%%%%%%

In Fig.~\ref{fig:Fresnel} we show an example of the vector potential  
in the time domain for various observers in the near field region of the 
shower as obtained in full ZHS simulations. The shower has an energy $E=100$ PeV and a longitudinal dimension 
of  $\sim 25$ m, and the observer is placed at different positions 
$z$ along the shower axis and at a fixed radial distance $r=10$ m.   
The ZHS simulation also gives the longitudinal profile of the excess
charge $Q(z')$ which we have introduced into Eq.~(\ref{eq:model_Fres})
to obtain the vector potential and compare it to that obtained directly 
by the Monte Carlo, also shown in Fig.~\ref{fig:Fresnel}. The agreement between the 
vector potential obtained directly in the Monte Carlo simulation 
and the calculation in our approach is remarkable. 
For distances to the shower axis larger than $\sim1~{\rm m}$ the
difference between our approach and the Monte Carlo is typically $\sim
1~\%$ or better for values down to $\sim 3$ orders of magnitude below
the peak of the vector potential.  
This difference starts to increase gradually as the distance to the shower axis 
decreases and becomes comparable to the lateral
dimensions of the shower, where the parameterization of the vector
potential at the Cherenkov angle given in Eq.~(\ref{eq:vp_fit}) 
is not expected to be valid. 
Since the typical distance between antennas in experiments
such as the Askaryan Radio Array (ARA) \cite{ARA} is $\sim 10 - 100$ m,
we expect our results to be accurate enough in most practical situations.

%%%%%%%%%%%%%%%%%%%%%%%%%%%%%%%%%%%%%%%%%%%%%%%%%%%%%%%%%%%%%%%%%%%%%%%%%%%%%%
\section{The apparent motion of a charge distribution.}

The temporal behavior of the vector potential traces the motion of a charged particle according to the retarded time. 
This is an old idea discussed by Feynman in \cite{Feynman1963} applied to elucidate on the properties of synchrotron radiation. More recently, this approach has also been used in the Cherenkov radiation calculation due to linear tracks in the near field~\cite{Afanasiev_book} and one dimensional current densities in~\cite{Chen_ARENA_2010}.
In this section we analyze the characteristics of our results in terms of the apparent motion of the charge density distribution to gain an intuitive understanding of the radiation due to a particle shower developing in a homogeneous dielectric medium. 

\subsection{Time delay effects}

The apparent motion of particles is encoded in the Green's function solution of the vector potential, Eq.~(\ref{eq:vp}), by the argument of the delta function taking the source time $t'$ to the observer time $t$: 
\begin{equation}
t = t' + \frac{n |\mathbf{x}-\mathbf{x}'|}{c}
\end{equation}
The term $|\mathbf{x}-\mathbf{x}'|$ traces the 
motion of the current density vector $\mathbf{J}$ at position $\mathbf{x}'(t')$ to determine the observer time. 

We can gain much insight into the properties of the vector potential resulting from the ZHS particle shower simulation, shown in Fig.~\ref{fig:Fresnel}, by momentarily ignoring the lateral distribution of the shower. 
In this case the observer time $t$ is given in terms of the source time $t'$ by 
\begin{equation}
t=t'+\frac{n\sqrt{r^2+(z-vt')^2}}{c}
\label{eq:times}
\end{equation}
where we have substituted $z'=vt'$ and $r$ is the cylindrical radial position of the observer. 

%%%%%%%%%%%%%%%%%%%%%%%%%%%%%%%%%%%%%%%%%%
\begin{figure}[tb]
\begin{center}
\includegraphics[width=\linewidth]{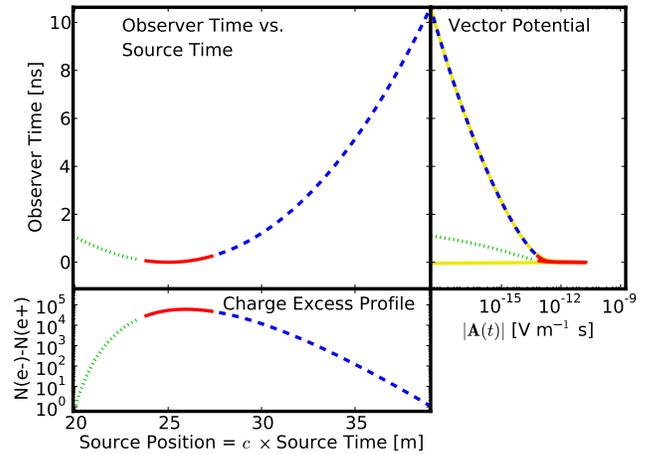}
\caption{Illustration of the vector potential characteristics due to the apparent motion of the charge distribution
for an observer seeing the region around shower maximum with angles close to the Cherenkov angle in ice ($n=1.78$). 
In the bottom plot we show the longitudinal charge excess profile following a Greisen parameterization.
The center plot shows the observer time as a function of the source time for a shower exceeding the speed of light 
in a medium. The solid (red) portion, corresponding to the minimum of the source and observer time relation, 
contributes significantly to the radiation near the Cherenkov peak.
The resulting vector potential is shown in the right hand side. 
The Cherenkov radiation spike corresponds to the compressed mapping of the charge excess distribution to the observer time.}
\label{fig:apparent}
\end{center}
\end{figure}
%%%%%%%%%%%%%%%%%%%%%%%%%%%%%%%%%%%%%%%%%%

When $v<c/n$, Eq. (\ref{eq:times}) has a unique observer time corresponding to each source time. However, in the case of $v>c/n$ 
there always exist a range of observer positions such that
Eq. (\ref{eq:times}) has two source times corresponding to every observer time (see an example in 
the top left panel of Fig.~\ref{fig:apparent}). 
In addition, a minimum value of $t$ (different from the trivial minimum $z'_0$ corresponding 
the beginning of the shower at $t'_0$) 
may exist when $n\beta>1$. In other words, the 
observer first sees the radiation corresponding to a depth in the shower 
$z'_{\mbox{\footnotesize{min}}}\neq z'_0$ and then sees contributions from shower depths before and 
after arriving simultaneously. 
This minimum can be characterized by looking at the derivative of the retarded time relation,
\begin{equation}
\frac{\partial t}{\partial t'} = 1-n\beta\frac{z-vt'}{\sqrt{r^2+(z-vt')^2}}
\end{equation}
The extrema in the relation between the source time and the observer
time are given by requiring 
the above equation to be equal to zero. A solution exists only when $v>c/n$  
and indeed corresponds to a minimum value 
in the observer time $t_{\mbox{\footnotesize{min}}}$.
The corresponding shower coordinate $z'_{\mbox{\footnotesize{min}}}$ given by the source time 
$t'_{\mbox{\footnotesize{min}}}=z'_{\mbox{\footnotesize{min}}}/v$ is 
\begin{equation}
z'_{\mbox{\footnotesize{min}}}=z+\frac{r}{\sqrt{n^2\beta^2-1}}.
\label{eq:zp_min}
\end{equation}

The angle of observation $\theta$ with respect to the shower axis corresponding to the shower position $z'$ is given by:
\begin{equation}
\cos\theta = \frac{(z-z')}{\sqrt{r^2+(z-z')^2}}
\label{eq:cher}
\end{equation} 
when $z'=z'_{\mbox{\footnotesize{min}}}$ it is straightforward to show that the angle corresponds to 
the Cherenkov angle $\cos\theta_C=1/(n\beta)$, i.e.
the minimum time $t_{\mbox{\footnotesize{min}}}$ at which the observer first sees the shower corresponds to the shower coordinate lying at the Cherenkov angle. Its value is given by
\begin{equation}
t_{\mbox{\footnotesize{min}}}=\frac{z}{v}+\frac{r\sqrt{n^2\beta^2-1}}{v}
\label{eq:t_min}
\end{equation}
A peculiar consequence of these relations is that if an observer is placed at a position such that the shower is always seen with $\theta<\theta_C$ then $z'_{\rm min}$ corresponds to the
end of the shower, which is an apparent violation of causality. In the case where $\theta>\theta_C$ then $z'_{\rm min}$ corresponds to the beginning of the shower as expected. This relation can be seen in Fig.~\ref{fig:times-angles} and is discussed in depth later in this section.

In the analytical solution (Fig.~\ref{fig:apparent}), for that particular observer seeing
the region around shower maximum with angles close to the Cherenkov angle, it is evident 
that a given observation time $t$ 
corresponds to two different shower coordinates $z'_\pm$ one
corresponding to an early development of the shower observed at angle $\theta<\theta_C$ and the other to 
a late one at angle $\theta>\theta_C$. When viewing particle showers
around the Cherenkov angle in the near-field, 
the radiation due to the early parts of the shower interferes with radiation due to the late parts of the shower. 
The apparent violation of causality is a relativistic effect due to the index
of refraction of the medium being $n>1$. Note also that for shower positions observed below the Cherenkov angle 
the derivative $\partial t / \partial t' <0$ meaning that time appears to run backwards. 

The depths $z'_\pm$ are 
easily obtained by expressing the source time $t'$ in terms of the observer time $t$ by inverting
Eq.~(\ref{eq:times}): 
\begin{equation}
t'_\pm=\frac{z'_\pm}{v}=\frac{vz-c^2_n t}{(v^2-c_n^2)}\pm c_n\frac{\sqrt{(z-vt)^2-r^2(n^2\beta^2-1)}}{(v^2-c_n^2)}
\end{equation}
where $c_n=c/n$. Real solutions exist if the argument of the square root is
non-negative which is equivalent to $t>t_{\mbox{\footnotesize{min}}}$ with $t_{\mbox{\footnotesize{min}}}$ 
given in Eq.~(\ref{eq:t_min}).

%%%%%%%%%%%%%%%%%%%%%%%%%%%%%%%%%%%%%%%%%%%%%%

The features of the vector potential due to the apparent motion of a charge distribution along an axis with $v>c/n$  are illustrated in Fig.~\ref{fig:apparent}. In the bottom panel we show 
the Greisen parametrized longitudinal
profile of the excess charge as a function of $z'$ or equivalently $t'$. In the 
center panel we show the relation between $t$ and $t'$ given in 
Eq.~(\ref{eq:times}) for ice ($n=1.78)$ with the shower front traveling at the speed of light. This relation has a minimum at $t'_{\mbox{\footnotesize{min}}}=z'_{\mbox{\footnotesize{min}}}/v$
given in Eq.~(\ref{eq:zp_min}) which corresponds to an observer time $t_{\mbox{\footnotesize{min}}}$
given in Eq.~(\ref{eq:t_min}). Around $t'_{\mbox{\footnotesize{min}}}$ the derivative $\partial t/\partial t'$
is very small and as a consequence the charge density of the source that corresponds to a relatively
wide interval of source times $t'_{\mbox{\footnotesize{min}}}\pm\Delta t'=t'_{\mbox{\footnotesize{min}}}\pm\Delta z'/v$, is projected and seen by the observer
during a small interval $t_{\mbox{\footnotesize{min}}}\pm\Delta t$ where,
\begin{equation}
\frac{\Delta t}{\Delta t'} = \frac{\partial t}{\partial t'}
\end{equation} 
This causes a time compression which enhances the radiation, especially when the geometry is such
that at $t_{\mbox{\footnotesize{min}}}$ the observer located at $(r,0,z)$ sees the region around the shower maximum.
This is illustrated in the right panel of Fig.~\ref{fig:apparent} where the vector potential corresponding
to the charge distribution shown in the bottom panel is depicted. The sharp initial peak of the
vector potential is due to the compression of radiation into a short interval of time as seen by
the observer.
For late enough observer times, the relation between the apparent and the observer time approaches linearity and the relation is causal. The observer sees later parts of the shower, shown in the bottom
panel of Fig.~\ref{fig:apparent}, at later times. 
This corresponds to the long tail in the vector potential depicted in the right panel of Fig.~\ref{fig:apparent}.
It is worth remarking
that the relativistic effects described here arise in any situation
where $v > c/n$ and are not limited to the description of Askaryan
radiation \cite{zhaires_air11}.

\subsection{Interpretation of simulation results}

\begin{figure}[tb]
\begin{center}
\includegraphics[width=\linewidth]{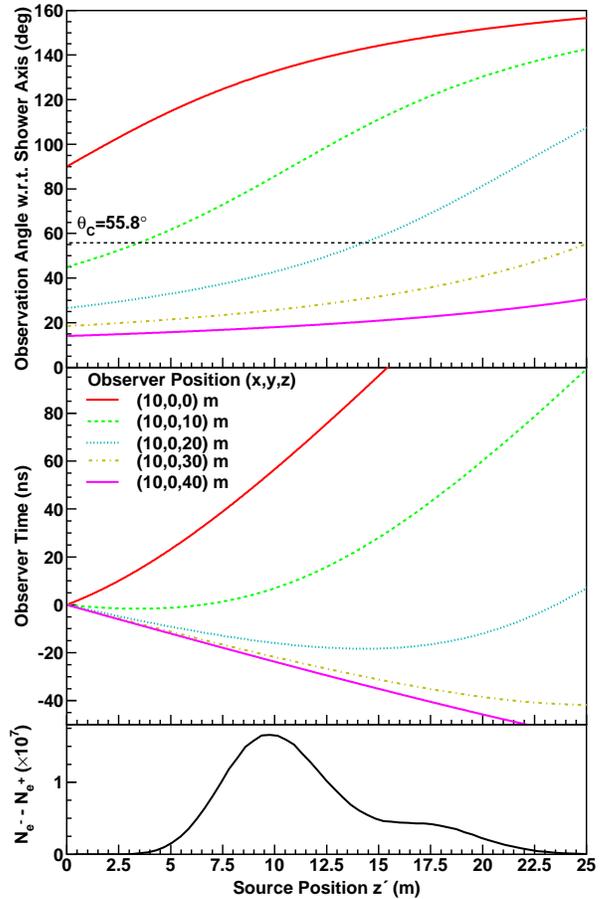}
\caption{Apparent motion analysis of the charge distribution for the vector potentials shown in Fig.~\ref{fig:Fresnel}. Bottom panel: Longitudinal profile of the excess negative charge in the 100 PeV electron-induced shower. Middle panel: Observer
time $t$ vs source position $z'$ for the same observers as in Fig~\ref{fig:Fresnel}.
Top panel: Angle with which each observer sees the different positions $z'$ 
in the shower axis. The Cherenkov angle ($\theta_C=55.8^\circ$ in ice) is indicated with a long-dashed horizontal line. See text for a description of the mapping of these relations to the features of the vector potentials shown in Fig.~\ref{fig:Fresnel}.}
\label{fig:times-angles}
\end{center}
\end{figure}

The corresponding time delay analysis of the results of the ZHS particle simulation shown in Fig.~\ref{fig:Fresnel} 
is displayed in Fig.~\ref{fig:times-angles}, using the same profile and source positions as 
depicted in Fig.~\ref{fig:Fresnel}. The observer located at $(x,y,z)=(10,0,20)$~m  
in  Figs.~\ref{fig:Fresnel} and \ref{fig:times-angles}, 
sees a sharp and strong spike in the vector potential (and electric field) that is not matched by that seen by observers
located at other positions in Fig.~\ref{fig:Fresnel}. 
In the middle panel of Fig.~\ref{fig:times-angles} we show the corresponding observer time $t$ 
vs. the source position $z'$ relation. 
For the observer at $(10,0,20)$~m one can clearly see a region with a small derivative 
$\partial t/\partial t'$ responsible for the compression effect which leads to an enhancement
in the vector potential. 
As shown with the aid of Eq.~(\ref{eq:cher}) the time at which an observer sees the shower 
first corresponds to observation at the Cherenkov angle. This can be also seen 
in the top panel of Fig.~\ref{fig:times-angles} 
where we have plotted the angle between the position $z'$ along the shower axis and the location 
of the observer. 
An observer at $(10,0,10)$~m also sees a fraction 
of the shower with angles around the Cherenkov angle. The Cherenkov pulse is not as pronounced because the net charge in that region is significantly smaller than the charge at shower maximum, 
and does not last as long as what the observer at $(10,0,20)$~m sees. 
In this view Cherenkov radiation is a geometrical phenomenon due to the minimum in the relation between the observer time 
and the source position\footnote{One cannot forget that the geometrical effect manifested as an index of refraction $n>1$ is in fact due to an interaction of the excess charge with the atoms in the medium and which is, in general, frequency dependent. The discussion in this section is only relevant to frequency bands where the index of refraction is reasonably approximated by a constant.}.

%%%%%%%%%%%%%%%%%%%%%%%%%%%%%%%%%%%%%%%%%%
%%%%%%%%%%%%%%%%%%%%%%%%%%%%%%%%%%%%%%%%%%

The apparent causality violations are manifested in the shape of the vector potential as viewed by different observers. 
An observer at an angle smaller than the Cherenkov
angle will see the evolution of the shower with an inverted causal order at all times.
This is the case of the vector potential labeled  $(x,y,z)=(10,0,40)$~m in Fig.~\ref{fig:Fresnel} where 
the observer sees a $\sim 25$ m long 100 PeV energy shower (shown in the inset of Fig.~\ref{fig:Fresnel} and in the 
bottom of Fig.~\ref{fig:times-angles})
 with the radiation corresponding to larger depths at later times. 
The longitudinal charge excess distribution shown in Fig.~\ref{fig:times-angles} has a primary peak followed 
by a smaller secondary peak. 
The vector potential traces this feature but in the reversed time sequence.
In the case of an observation at angles larger than $\theta_C$, the shower is seen 
in the normal causal order at all times. This case is illustrated by vector potential labeled  $(x,y,z)=(10,0,10)$ m in 
Fig.~\ref{fig:Fresnel} where the radiation corresponding to the charge excess distribution at larger depths 
is observed at later times. 
The peaks of the vector potential match the order of the peaks of the charge excess distribution in the bottom 
of Fig.~\ref{fig:times-angles}. 
It is also worth noting that even though single antenna measurements in the near-field make it difficult to 
reconstruct the longitudinal profile of the charge excess, observations from multiple stations 
located at tens of meters from the shower axis, do provide the necessary information.

%%%%%%%%%%%%%%%%%%%%%%%%%%%%%%%%%%%%%%%%%%
\section{Summary and Outlook}

We have derived a highly detailed and computationally efficient approach for the 
calculation of Askaryan pulses. 
The electrodynamic calculations leading to the relation between the pulse features and 
the shower characteristics can be intuitively understood via the apparent motion of charges.
Viewed from this perspective, one can easily retrace the time-domain
 behavior of the pulse to the shape of the electromagnetic shower and the observer versus source time relation as 
shown in Fig.~\ref{fig:apparent}.   

There are many interesting features of Askaryan radiation due to the fact that the speed of the shower front exceeds
 the speed of light. In this treatment, the radiation due to the charge excess of an electromagnetic shower 
is understood as a ``dense" or compressed mapping of the charge excess profile to the vector potential via the observer 
vs. source time relation. The mapping is densest at the minimum of the observer vs. source time relation which 
corresponds to observations at the Cherenkov angle. In addition, for observation at angles smaller than the Cherenkov angle, time 
appears to run backwards. This manifests itself in the time-reversed mapping of the longitudinal profile of the charge 
excess to the time-domain vector potential.
 
The primary motivation for developing this calculation in the time-domain is to 
understand the temporal behavior of the Askaryan pulse. In the frequency domain, this 
is equivalent to understanding the phase versus frequency relation. Although it is 
possible to do this in a completely frequency-domain approach, the time-domain 
relations can be intuitively understood and are easier to compute. 

The computational algorithm presented here is summarized as follows: 

\begin{enumerate}

\item 
Compute the vector potential of the Askaryan pulse at the Cherenkov
angle $\mathbf{A}(\theta_C,t)$ and use it together with total charged track-length  
$LQ_{tot}=\int dz'Q(z')$  to extract the functional form of the form factor $F_p$. 
(This has been done in this article for electromagnetic showers in ice
and in that case it is possible to use directly the parameterization provided in 
Eq.~(\ref{eq:vp_fit}). For other situations it needs to be
re-evaluated with a detailed Monte Carlo simulation.)

\item 
Obtain the charge excess longitudinal profile of an electromagnetic
shower~$Q(z')$. This can be provided as either the output of a particle shower
simulation or using a parameterization. 

\item
Convolve $F_p$ with $Q(z')$ according to Eq.~(\ref{eq:model_Fraun})
in the far-field or Eq.~(\ref{eq:model_Fres}) in the near-field to obtain the time-domain vector potential. 

\item 
Electric fields are obtained from a trivial numerical derivative of the vector potential with respect to time:
${\mathbf{E}}=-\partial {\mathbf{A}}/\partial t$. 
\end{enumerate}

The formalism developed here can also be applied to the reconstruction
of longitudinal shower profiles. In the far-field this can only be
done if the pulse was detected away from the Cherenkov angle,
otherwise the pulse shape is approximately the same for any given
longitudinal charge excess profile. 
Away from the Cherenkov angle the pulse traces the shape of the
longitudinal profile convolved with the lateral profile response. 
For very extended longitudinal profiles, 
such as those resulting from UHE showers affected by the LPM effect,
the tracing of the profile can be seen for angles as 
small as $0.3^\circ$ away from the Cherenkov angle (see Fig.~\ref{fig:LPM}). 
The overall quality of the reconstruction 
can be assessed with simulations that are specific to the experiment in question. 
The reconstruction of longitudinal profiles has interesting experimental applications 
such as the identification of the primary particle or $\nu$ flavor inducing the shower. 
This is particularly relevant for the electromagnetic component of a $\nu_e$-induced 
shower with its multiply-peaked structure due to the LPM effect.

In the near-field the reconstruction of shower longitudinal profiles is also possible. 
If the radiation is detected by a single station the reconstruction is
complicated by the fact that the portions of the shower above and below the Cherenkov 
angle may interfere with each other depending on the position of the antenna. However, 
if multiple antennas observe the radiation due to a 
single shower it is possible to obtain a highly constrained
reconstruction of the longitudinal profile. The example 
depicted in Fig.~\ref{fig:Fresnel} shows a case where this could be
done for antennas in ice spaced tens of meters apart. 
The formalism provided in this paper will allow the experimentalist to simulate these measurements and find the 
optimal antenna placement and assess the quality of reconstruction. This would be of particular interest to an 
array such as the future planned ARA \cite{ARA} and ARIANNA \cite{ARIANNA} experiments where the shower could 
potentially be observed by multiple stations in Antarctic ice.
 
In a future publication we plan to produce time-domain parameterizations of the vector potential at the Cherenkov angle for electromagnetic showers in various media such as salt and the lunar regolith. 
In addition the time-domain parameterizations of hadronic showers will be included which will allow the experimentalist to produce 
full simulations of neutrino interactions with flavor dependent parameters. This will be useful for experiment simulations and candidate event reconstructions.

\section{Acknowledgments}
J.A-M and E.Z. thank Xunta de Galicia (INCITE09 206 336 PR) and 
Conseller\'\i a de Educaci\'on (Grupos de Referencia Competitivos -- 
Consolider Xunta de Galicia 2006/51); Ministerio de Ciencia e Innovaci\'on 
(FPA 2007-65114, FPA 2008-01177 and Consolider CPAN - Ingenio 2010) and Feder Funds, Spain.
We thank CESGA (Centro de SuperComputaci\'on de Galicia) for computing resources
and assistance. A.R-W thanks NASA (NESSF Grant NNX07AO05H). Part of this research was carried out at the Jet
Propulsion Laboratory, California Institute of Technology, under a contract with the National Aeronautics and Space Administration.
We thank J. Bray, P. Gorham, C.W. James and W.R. Carvalho Jr. for helpful discussions.

\end{document}